\documentclass[prd,aps,amsmath,amssymb,epsfig]{revtex4}

\usepackage{graphicx}
\usepackage{dcolumn}
\usepackage{bm}
\usepackage{feynmp}

\newcommand{\bea}{\begin{eqnarray}}
\newcommand{\eea}{\end{eqnarray}}
\begin{document}
\title{Spontaneous symmetry breaking in gauge theories via Bose-Einstein
condensation}
 \author{Francesco {\sc Sannino}}
 \email{francesco.sannino@nbi.dk}
 \author{Kimmo {\sc Tuominen}}\email{tuominen@nordita.dk}
 \affiliation{{\rm NORDITA}, Blegdamsvej 17, DK-2100 Copenhagen \O, Denmark }
\date{March 2002}

\begin{abstract}
We propose a mechanism naturally leading to the spontaneous
symmetry breaking in a gauge theory. The Higgs field is assumed to
have global and gauged internal symmetries. We associate a non
zero chemical potential to one of the globally conserved charges
commuting with all of the gauge transformations. This induces a
negative mass squared for the Higgs field triggering the
spontaneous symmetry breaking of the global and local symmetries.
The mechanism is general and we test the idea for the electroweak
theory in which the Higgs sector is extended to possess an extra
global Abelian symmetry. To this symmetry we associate a non zero
chemical potential. The Bose-Einstein condensation of the Higgs
leads, at tree level, to modified dispersion relations for the
Higgs field while the dispersion relations of the gauge bosons and
fermions remain undisturbed. The latter are modified through
higher order corrections. We have computed some corrections to the
vacuum polarizations of the gauge bosons and fermions. To quantify
the corrections to the gauge boson vacuum polarizations with
respect to the Standard Model we considered the effects on the $T$
parameter. We finally derive the one loop modified fermion
dispersion relations.
\end{abstract}

\maketitle

\section{Introduction}
\label{introduction}

The Standard Model of particle interactions has been much studied
and has passed numerous experimental tests \cite{Hagiwara:fs}.
Despite its experimental successes it is commonly believed that it
is yet an incomplete model. Within the Standard Model, for
example, one does not know what controls the spontaneous breaking
of the electroweak symmetry since the negative mass squared of the
Higgs particle is simply assumed. Different models have been
proposed to address the problem of electroweak symmetry breaking.
Technicolor theories \cite{Hill:2002ap} and supersymmetric
extensions \cite{Kane:2002tr} of the Standard Model are two
relevant examples.

In this paper we explore the possibility that the relativistic
Bose-Einstein phenomenon, much studied in the literature
\cite{Haber:1981ts,Kapusta:aa}, is the source of spontaneous
symmetry breaking of a generic gauge theory. We introduce a
chemical potential $\mu$ for some global symmetries of the Higgs
field. The generators associated to these symmetries are chosen to
commute with the gauge transformations. A relevant property of the
theory is that the chemical potential induces a negative mass
squared for the Higgs field at the tree level hence destabilizing
the symmetric vacuum and triggering symmetry breaking. The local
gauge symmetries are broken spontaneously and the associated gauge
bosons acquire a standard mass term. This is so since we have
chosen the global generator associated with the chemical potential
to commute with the gauge transformations. In this way the mass of
the Higgs boson is intrinsically related to a given charge density
while its potential is in fact (at zero temperature) the
thermodynamical potential. While the properties of the massive
gauge bosons at the tree level are identical to the ones induced
by the conventional Higgs mechanism, the Higgs field encodes all
of the information associated to this different way of symmetry
breaking. It has specific non Lorentz covariant dispersion
relations since the introduction of the chemical potential
differentiates between space and time. However the directly
observable part of the theory feels Lorentz breaking effects via
radiative corrections involving the Higgs field. The size of the
corrections to the dispersion relations of a given particle is
controlled by the relative coupling strength of the various
particles with the Higgs.

We start our analysis in a pedagogical way by first reviewing how
spontaneous symmetry breaking manifests itself in a relativistic
non ideal (i.e. self-interacting) bosonic $U(1)$ invariant theory
with an associated non zero chemical potential. Via this simple
model we familiarize with the scalar field dispersion relations in
the spontaneously broken phase. We then consider an example of a
gauge theory in which spontaneous symmetry breaking is entirely
due to the introduction of a specific chemical potential. The
gauge symmetry is taken to be identical to the one of the Standard
Model i.e. $SU_L(2)\times U_Y(1)$.

The Higgs sector of the theory is minimally extended to enable the
inclusion of a chemical potential not associated with any of the
gauge generators. In the Standard Model the Higgs sector has an
accidental $SU_L(2)\times SU_R(2)$ global symmetry where the
$SU_L(2)$ is gauged and the hypercharge is associated to the
diagonal $SU_R(2)$ generator. We consider the extension where the
Higgs sector contains an extra $U_A(1)$ symmetry. To this extra
global symmetry we then associate a non zero chemical potential
which enters with a negative mass squared term in the Lagrangian
for the Higgs leading to the destabilization of the vacuum and
spontaneous symmetry breaking. So the $U_A(1)\times SU_L(2)\times
SU_R(2)$ breaks to an unbroken $SU_V(2)$ with three of the gauge
bosons acquiring a mass term while the photon remains massless.
$SU_V(2)$ is the custodial symmetry which at the tree level
guarantees the relation $M^2_W=M^2_Z\,\cos^2\theta_W$, where $M_W$
is the mass of the electromagnetically charged gauge bosons, $M_Z$
is the $Z$ neutral gauge boson mass and $\theta_W$ is the Weinberg
angle. {}Three of the four goldstone bosons, in the unitary gauge,
become the longitudinal components of the massive gauge bosons
while we are left with an extra massless (i.e. gapless) degree of
freedom $\eta$. In the theory we have one more singlet field which
is the neutral Higgs $h$. Due to the presence of the chemical
potential a time dependent Lorentz breaking mixing term between
the $h$ and the $\eta$ field emerges. This is the term responsible
for non conventional dispersion relations for the $h$ and $\eta$
field. The $\eta$ gap (the energy at zero momentum) is still zero
while the $h$ gap is slightly higher than the mass associated to
the potential curvature evaluated on the vacuum of the theory.
Hence we predict a Higgs sector with very specific dispersion
relations and mass spectrum. Different models where the $\eta$
field is not light can also be constructed.

All of the gauge bosons, at the tree level, posses standard
Lorentz covariant dispersion relations and the Lorentz breaking
effects for those are felt only via radiative corrections. The
diagrams inducing Lorentz breaking include the exchange of the
Higgs and/or $\eta$ particles. We have computed the relevant zero
momentum one loop contribution to the $W$ and $Z$ vacuum
polarizations. The details of the computation are in the appendix.
Since the chemical potential differentiates between time and space
and we have a scalar vacuum the dispersion relations are
isotropic. Theories with condensates of vectorial type have been
studied in different realms of theoretical physics
\cite{{Linde:1979pr},{Ferrer:jc},{Ferrer:ag},{Ambjorn:1989gb},{Kajantie:1998rz},{Sannino:2002wp},{Manton:1979kb},{Hosotani:1988bm},{Li:2002iw}}.
To quantify the corrections with respect to the Standard Model, we
have generalized the oblique parameter $T$
\cite{Peskin:1991sw,Takeuchi:1992bu} to have a Lorentz structure
$T^{\mu \nu}\propto \left({\Pi_{11}^{\mu\nu}}^{new}(0) -
{\Pi_{33}^{\mu\nu}}^{new}(0)\right)$ where $\Pi^{new}$ indicates
the new physics contributions to the vacuum polarization of the
vector bosons. The Lorentz breaking sector preserves the custodial
symmetry and we find $T^{\mu\nu}=T\,g^{\mu\nu}$. However there is
an effect on $T$ due to the fact that the $h$ gap is different
than the ordinary mass in the Standard Model yielding $T\simeq
-0.048$. The fermionic sector of the Standard Model has also been
investigated. We have shown that for a generic fermion the
dispersion relations are $E^2 = v_{f}^2\,p^2+m_f^2$. At the one
loop level due to the smallness of the Yukawa coupling $v_f$ is
small.

Effects of a large lepton number on the spontaneous gauge symmetry
breaking for the electroweak theory at high temperature relevant
for the early universe have been studied in the literature
\cite{Linde:1979pr,{Ferrer:jc},{Ferrer:ag},{Bajc:1999he},{Bajc:1997ky}}.

The issue of Lorentz breaking has attracted recently theoretical
\cite{Kostelecky:2002hh,Carroll:2001ws,{Amelino-Camelia:2002dx}}
and experimental attention \cite{Carroll:vb}. For example in
\cite{Carroll:2001ws} possible extensions of the Standard Model
were studied in which Lorentz symmetry is violated. In
\cite{Kostelecky:2002hh,Carroll:2001ws} the authors investigate
possible phenomenological consequences due to explicit Lorentz
breaking terms added to the Standard Model Lagrangian. In the
present case Lorentz breaking is due to a net background charge
density.

In section \ref{due} we briefly summarize the basic properties of
Bose-Einstein condensation relevant for our theory. We investigate
spontaneous symmetry breaking in a gauge theory due to
Bose-Einstein condensation in section \ref{tre}. The electroweak
gauge sector of the Standard Model is chosen as a fundamental
example to test our idea. We analyze the tree level and some of
the higher order corrections in section \ref{quattro}. We finally
compute the general structure of the corrections to the fermion
dispersion relations and predict the size of the corrections
within our framework. Experiments studying the dispersion
relations for different fermions may be able to test our
predictions. Finally we summarize in \ref{cinque} while suggesting
different possible investigations and/or extensions. An appendix
in which we provide some useful results and explicit evaluations
of diagrams concludes the paper.

\section{Bose-Einstein Condensation in a nutshell}\label{due}
The effects of a net background charge on ideal and interacting
relativistic Bose gases with Abelian and non Abelian symmetries
have been investigated in the literature
\cite{Haber:1981ts,Kapusta:aa}. In this section we briefly review
the basic aspects related to the introduction of a net background
charge for a bosonic theory with $U(1)$ invariance which we will
then use when extending the theory to describe the electroweak
symmetry breaking sector of the Standard Model. It has been shown
that the chemical potential associated to a net background charge
for a charged scalar field must be introduced at the Lagrangian
level as follows\cite{Haber:1981ts,{Kapusta:aa}}:
\begin{eqnarray}{\cal L}=D_{\mu}\phi^{\ast} D^{\mu}\phi -m^2
|\phi|^2 -\frac{\lambda}{4}|\phi|^4 \ , \label{bose}
\end{eqnarray}
with $m^2$ and $\lambda$ positive constants and
\begin{eqnarray}
D_{\nu} \phi=\partial_{\nu}\phi - i A_{\nu}\phi \ , \qquad A_{\nu}=\mu
\left(1,\vec{0}\right) \ , \label{covariant}
\end{eqnarray}
and $\mu$ the associated chemical potential. Substituting eq.(\ref{covariant}) in
eq.({\ref{bose}}) we have:
\begin{eqnarray}
{\cal L}=\partial_{\mu}\phi^{\ast}\partial^{\mu}\phi
+i\,\mu\left(\phi^{\ast}\partial_{0}\phi -
\partial_{0}\phi^{\ast}\phi\right)-\left(m^2 - \mu^2\right)|\phi|^2
-\frac{\lambda}{4}|\phi|^4 \ .\end{eqnarray} The grand canonical partition function is:
\begin{eqnarray}Z=\int[d\phi]\,[d\phi^{\ast}]\,\exp\left[\int^\beta_0 \,d\tau \int_V\,d^3x\, {\cal L} \right]\ ,\end{eqnarray}
where $\tau$ is euclidian time and $\beta=1/T$ is the inverse of
the temperature. We will be working in the following at zero
temperature. The introduction of the chemical potential has broken
Lorentz invariance $SO(1,3)$ to $SO(3)$ while providing a {\it
negative} mass squared contribution to the boson. This is at the
heart of the Bose-Einstein condensation phenomenon. When $\mu>m$
we have spontaneous breaking of $U(1)$ invariance. We will later
on exploit this basic feature to provide an alternative way for
spontaneously breaking a local rather than global internal
symmetry. Defining
\begin{eqnarray}
\phi=\langle \phi \rangle + \frac{1}{\sqrt{2}} \left(h + i\,\eta \right)\ ,
\end{eqnarray}
the quadratic term in the fields reads:
\begin{eqnarray}
L_{quadratic}=\frac{1}{2} \left[ h \, \,\,\, \eta \right] \, \left[%
\begin{array}{cc}
  -\partial^2 -(m^2 - \mu^2) -\frac{3\lambda}{2}\langle \phi \rangle^2 & -2\,\mu\,\partial_{0} \\
  2\,\mu\,\partial_0 & -\partial^2 -(m^2 - \mu^2)-\frac{\lambda}{2}\langle \phi \rangle^2   \\
\end{array}%
\right] \, \left[%
\begin{array}{c}
  h \\
  \eta \\
\end{array}%
\right]
\end{eqnarray}
where we have chosen $\langle \phi \rangle$ real. $\langle \phi \rangle = 0$ for $\mu\leq
m$ while
\begin{eqnarray}
\langle \phi \rangle^2 = \frac{2}{\lambda}\,\left(\mu^2 - m^2\right) \ ,
\end{eqnarray}
for $\mu>m$. In the broken phase we have a gapless (i.e. the energy at zero momentum)
excitation as well as a gapped excitation. The kinetic term matrix in the momentum space
is:
\begin{eqnarray}
L_{quadratic}&=&\frac{1}{2} \left[ h (-p)\, , \,\eta (-p)\right] \, D(p)\,\left[%
\begin{array}{c}
  h (p)\\
  \eta (p) \\
\end{array}%
\right]  \ , \\ &&\nonumber \\ && \nonumber \\
D(p)&=&\left[%
\begin{array}{cc}
  p^2_0 - {\bf p}^2-(m^2 - \mu^2) -\frac{3\lambda}{2}\langle \phi \rangle^2 & 2\,i\,\mu\,p_{0} \\
  -2\,i\,\mu\,p_0 & p^2_0 -{\bf p}^2 -(m^2 - \mu^2)-\frac{\lambda}{2}\langle \phi \rangle^2   \\
\end{array}%
\right] \ .
\end{eqnarray}
Imposing, as customary, the vanishing of the determinant of $D(p)$
one determines the tree level dispersion relations. In the broken
phase after expanding in the momentum we have:
\begin{eqnarray}
E^2_{h}&=&\Delta^2 + \left(1+4\frac{\mu^2}{\Delta^2}\right)\, p^2
- \frac{\mu^4}{\Delta^6}\, p^4 + \cdots \ ,\\
 E^2_{\eta}&=&
\left(1-4\frac{\mu^2}{\Delta^2}\right)\, p^2 + 16\,\frac{\mu^4}{\Delta^4}\, p^4 + \cdots \
,
\end{eqnarray}
with $\Delta^2 = 2\, \left(3\mu^2 - m^2\right)$. We note that for
$p_0$ and $p$ very large we recover the propagators of the free
theory for massless particles. So that all of the ultraviolet
properties of the theory are identical to the one before adding
the chemical potential at zero temperature. This observation
allows us immediately to conclude that if a theory were
renormalizable before adding the chemical potential it will remain
renormalizable after having introduced it. On the other hand for
small momenta and large $\mu$ we have a static field of mass
squared $\Delta^2= 6\,\mu^2$ and a massless state with
$E^2_{\eta}={\bf p}^2/3$ dispersion relation.
Rotational symmetry remains intact here. However theories in which
rotational symmetry breaks together with some internal global
symmetries due to the introduction of a chemical potential have
been studied in some detail in \cite{Sannino:2002wp}. The
propagator matrix is simply $i\, D(p)^{-1}$. It is instructive to
see, in some detail, how the $h$ decoupling works in the broken
phase. Indeed for large $\mu$ (setting $m=0$) the part of the
Lagrangian quadratic in the fields is:
\begin{eqnarray}
\frac{1}{2}\partial_{\mu}h\partial^{\mu}h
+\frac{1}{2}\partial_{\mu}\eta\partial^{\mu} \eta - 2\mu\,
h\partial_0\eta - 2\mu^2\, h^2 \ .
\end{eqnarray}
In the limit $\mu\rightarrow \infty$ we neglect the $h$ kinetic term and the Euler-Lagrange
equation for $h$ is:
\begin{eqnarray}
h=-\frac{1}{\mu} \partial_0 \eta \ .
\end{eqnarray}
Substituting this back into the previous expression leads to the following $\eta$
Lagrangian:
\begin{eqnarray}
\frac{1}{2}\left(3\,\partial_0\eta \partial_0\eta - \nabla\eta
\cdot \nabla \eta\right) \ . \label{etadispersions}\end{eqnarray}
Which leads correctly to the dispersion relation for $\eta$ at
small momenta with respect to the chemical potential, i.e.
$E^2_\eta = {\bf p}^2/3$. Note though that the $h$ gap is not the
curvature evaluated on the minimum which is $2\mu^2$ but is
$\Delta^2=6\mu^2$. Our discussion, clearly, does not depend on the
specific $U(1)$ charge assignment for $\phi$. However it is
important that the vev is in the $h$ direction.

At the tree level and at zero temperature the thermodynamic potential is simply:
\begin{eqnarray}
V[\langle \phi \rangle(\mu) , \mu]=\frac{\lambda}{4}\langle\phi \rangle^4 + \left(m^2
-\mu^2\right)\langle \phi \rangle^2 = -\frac{1}{\lambda}\left(\mu^2 - m^2 \right)^2\ .
\end{eqnarray}
In the standard Higgs mechanism the negative mass squared
parameter is not automatically linked to any underlying physical
quantity. In the Bose-Einstein case instead the chemical potential
driving spontaneous symmetry breaking is necessarily linked to the
following net charge density:
\begin{eqnarray}
\rho  = - \frac{\partial V[\langle \phi \rangle(\mu) , \mu]}{\partial \mu} =
\frac{4\mu}{\lambda} \left(\mu^2 - m^2\right) \ .
\end{eqnarray}
We now extend the internal symmetries of the present bosonic field
to the case in which some of the new internal symmetries are
gauged. We choose the gauged symmetries to be identical to the
ones associated to the electroweak sector of the Standard Model.
The spontaneous symmetry breaking in the gauge theory becomes
driven by the presence of a net background charge which manifests
itself via a non zero chemical potential.

\section{The $\mathbf{SU_L(2)\times U_Y(1)}$ Gauge
Theory}\label{tre} The Higgs sector of the Standard Model
possesses, when the gauge couplings are switched off, an
$SU_L(2)\times SU_R(2)$ symmetry. The full symmetry group is
mostly easily recognized when the Higgs
doublet field \begin{eqnarray} H=\frac{1}{\sqrt{2}}\left(%
\begin{array}{c}
  \pi_2 + i\, \pi_1 \\
  \sigma - i\, \pi_3 \\
\end{array}%
\right)\end{eqnarray} is represented as a two by two matrix in the
following way:
\begin{eqnarray}
\left[i\,\tau_2H^{\ast}\
,\,H\right]=\frac{1}{\sqrt{2}}\left(\sigma + i\,
\vec{\tau}\cdot\vec{\pi} \right) \equiv M
 \ .
\end{eqnarray}
The $SU_L(2)\times SU_R(2)$ group acts linearly on $M$ according
to:
\begin{eqnarray}
M\rightarrow g_L M g_R^{\dagger} \qquad {\rm and} \qquad g_{L/R} \in SU_{L/R}(2)\ .
\end{eqnarray}
It is easy to verify that:
\begin{eqnarray}
M\frac{\left(1-\tau^3\right)}{2} = \left[0\ , \, H\right] \ . \qquad
M\frac{\left(1+\tau^3\right)}{2} = \left[i\,\tau_2H^{\ast} \ , \, 0\right] \ .
\end{eqnarray}
The $SU_L(2)$ symmetry is gauged by introducing the weak gauge
bosons $W^a$ with $a=1,2,3$. The hypercharge generator is taken to
be the third generator of $SU_R(2)$. The ordinary covariant
derivative acting on the Higgs, in the present notation, is:
\begin{eqnarray}
D_{\mu}M=\partial_{\mu}M -i\,g\,W_{\mu}M + i\,g^{\prime}M\,B_{\mu} \ , \qquad {\rm
with}\qquad W_{\mu}=W_{\mu}^{a}\frac{\tau^{a}}{2} \ ,\quad
B_{\mu}=B_{\mu}\frac{\tau^{3}}{2} \ .
\end{eqnarray}
At this point one simply {\it assumes} that the mass squared of
the Higgs field is negative and this leads to the electroweak
symmetry breaking and more generally to the successful Standard
Model as we know. However, theoretically a more satisfactory
explanation of the origin of the Higgs mechanism is needed. In the
literature many models have been proposed in order to explain the
emergence of such a negative mass squared. Technicolor theories,
for example, assume a dynamical mechanism identical to spontaneous
chiral symmetry breaking in quantum chromodynamics
\cite{Hill:2002ap}. The strength of technicolor relies on the fact
that this mechanism has already been observed in nature. However
the simplest technicolor models lead to a too large $S$ parameter
\cite{Peskin:1991sw}. The $S$-parameter is not a problem in the
case of extended technicolor theories
\cite{{Appelquist:1998xf},Appelquist:1999dq} allowing for new
technicolor models \cite{Appelquist:2003uu}. Supersymmetric
extensions of the Standard Model \cite{Kane:2002tr} explain the
negative mass squared as due to the running of the masses from
high scales down to the electroweak one. The models mentioned
above are, at the moment, the generalizations of the Standard
Model able to explain the origin of electroweak symmetry breaking.

Here we propose a new mechanism based on the previously explained
Bose-Einstein phenomenon. To illustrate our idea we consider a
Higgs sector with the symmetry group $SU_L(2)\times SU_R(2)\times
U_A(1)$ where the $SU_L(2)$ is later on gauged and the $U_Y(1)$ is
associated to the $T^3=\frac{\tau^3}{2}$ generator of $SU_R(2)$
while $U_A(1)$ remains a global symmetry. Now we introduce a
chemical potential $\mu_A$ associated to $U_A(1)$. When the
chemical potential is sufficiently large $SU_L(2)\times
SU_R(2)\times U_A(1)$ breaks spontaneously to $SU_V(2)$ and we
have four goldstones. It is advantageous to use non linear
realizations for the Higgs field with:
\begin{eqnarray}
M=\frac{\sigma}{\sqrt{2}}\,U_{\eta}\,U \qquad {\rm with} \qquad
U_{\eta}=e^{i\frac{\eta}{v}} \,\qquad U=e^{i\frac{\pi}{v}}\ , \qquad {\rm and } \qquad
\pi={\tau^a}\pi^a \ .
\end{eqnarray}
In the above equation $v$ is the vacuum expectation value of
$\sigma$. In the linearly realized case we should also include the
heavy $U_A(1)$ partners of the $\pi$ field which we have taken to
be more massive than the neutral Higgs particle and hence we have
decoupled them. The $\eta$ field is the the goldstone boson
associated to the spontaneous breaking of the global $U_A(1)$
symmetry.

The gauge interactions as well as the chemical potential are introduced via the following
covariant derivative:
\begin{eqnarray}
{\cal D}_{\mu}M=\partial_{\mu}M - i\,g\,W_{\mu}\,M +
i\,g^{\prime}\,M\,B_{\mu}-i{\cal A}_{\mu}M \equiv {D}_{\mu}M -
i{\cal A}_{\mu}M\ , \quad {\rm with} \quad {\cal
A}_{\mu}={\mu_{A}} (1,\vec{0}) \ .
\end{eqnarray}
Substituting the covariant derivative in the generalized kinetic
term yields:
\begin{eqnarray}
{\rm Tr} \left[{\cal D}_{\mu}M^{\dagger} {\cal D}^{\mu}M\right]={\rm
Tr}\left[D_{\mu}M^{\dagger}D^{\mu}M\right] -i{\mu_A}{\rm Tr}\left[M D_0 M^{\dagger} -
M^{\dagger}D_{0}M\right] +{\mu_{A}^2}{\rm Tr}\left[M^{\dagger}M\right] \ .
\label{Derivative}
\end{eqnarray}
Electroweak breaking is now forced by the introduction of the
chemical potential for the extra global symmetry due to the
emergence of the third term in eq.~(\ref{Derivative}). Adding a
general Higgs potential the terms for the scalar sector of the
electroweak Lagrangian are:
\begin{eqnarray}
{\cal L}&=&\frac{1}{2}{\rm Tr} \left[D_{\mu}M^{\dagger}
D^{\mu}M\right] -i\frac{\mu_A}{2}{\rm Tr}\left[M D_0 M^{\dagger} -
M^{\dagger}D_{0}M\right] -\frac{1}{2}(m^2 -{\mu_{A}^2}){\rm
Tr}\left[M^{\dagger}M\right] - \frac{\lambda}{4}\,{\rm
Tr}\left[M^{\dagger}M\right]^2
\end{eqnarray}
{}For ${\mu_A^2}>m^2$ we have Bose-Einstein condensation together
with the ordinary spontaneous breaking of the internal symmetry
$SU_L(2)\times SU_R(2)\times U_A(1)\rightarrow SU_V(2)$  with 4
null curvatures corresponding to the four broken generators. In
the unitary gauge the three fields $\pi^a$ are absorbed into the
longitudinal components of the three massive gauge boson fields
while the field $\eta$ remains massless. In the unitary gauge the
quadratic terms are:
\begin{eqnarray}
{\cal L}_{\rm quadratic}&=&\frac{1}{2} \partial_{\mu}h
\partial^{\mu}h
+\frac{1}{2}\partial_{\mu}{\eta}\partial^{\mu}{\eta}  - \mu_A \left(h\partial_0 \eta - \eta
\partial_0 h\right) \nonumber \\&+&\frac{v^2}{8}\, \left[g^2\,\left(W_{\mu}^1
W^{\mu,1} +W_{\mu}^2 W^{\mu,2}\right)+ \left(g\,W_{\mu}^3 -
g^{\prime}\,B_{\mu}\right)^2\right] -\frac{1}{2}(\mu^2_A -{m^2})h^2 \ ,
\end{eqnarray}
with
\begin{eqnarray}\langle \sigma \rangle^2 =
v^2={\frac{{\mu_A^2}-m^2}{\lambda}}\ , \qquad {\rm and} \qquad \sigma = v + h \ ,
\end{eqnarray}
where $h$ is the Higgs field. The $Z_{\mu}$ and the photon $A_{\mu}$ gauge bosons are:
\begin{eqnarray}
Z_{\mu}&=&\cos\theta_W\, W_{\mu}^3 - \sin\theta_{W}B_{\mu} \ ,\nonumber \\
A_{\mu}&=&\cos\theta_W\, B_{\mu} + \sin\theta_{W}W_{\mu}^3 \ ,
\end{eqnarray}
with $\tan\theta_{W}=g^{\prime}/g$ while the charged massive vector bosons are
$W^{\pm}_{\mu}=(W^1\pm i\,W^2_{\mu})/\sqrt{2}$. The bosons masses $M^2_W=g^2\,v^2/4$ due to
the custodial symmetry satisfy the tree level relation $M^2_Z=M^2_W/\cos^2\theta_{W}$.

Except for the presence of an extra degree of freedom and the
$h-\eta$ mixing term one recovers the correct electroweak symmetry
breaking pattern. The third term in the Lagrangian signals an
explicit breaking of the Lorentz symmetry in the Higgs sector.
Such a breaking is due to the introduction of the chemical
potential and hence it happens in a very specific and predictive
way so that all of the features can be studied. Note that in our
theory Lorentz breaking, at the tree level, is confined only to
the Higgs sector of the theory which is also the least known
experimentally. The rest of the theory is affected via weak
radiative corrections. As previously emphasized, in general, the
introduction of the chemical potential at zero temperature does
not introduce new ultraviolet divergences and hence does not spoil
renormalizability. Besides, diagonalizing the quadratic terms the
spectrum and the propagators for $h$ and $\eta$ are identical to
the one presented in the first section with the replacement
$\mu\rightarrow \mu_A$, i.e.:
\begin{eqnarray}
E^2_{h}&=& \Delta^2 + \left(1 + 4\frac{\mu_{A}}{\Delta^2}\right)p^2 -
\frac{\mu_{A}^4}{\Delta^6}
p^4 + \cdots  \ , \\
E^2_{\eta}&=&\left(1 - 4\frac{\mu_{A}^2}{\Delta^2}\right)p^2 + \cdots.
\end{eqnarray}
with
\begin{eqnarray}
\Delta^2 = 2(3\mu^2_{A} - m^2)=4\mu^2_{A} + 2\,\left({\mu^2_{A}} -
m^2\right) \ .
\end{eqnarray}
The second term in the expression for $\Delta^2$ is the potential
curvature evaluated on the ground state which in the absence of
the chemical potential is the mass of $h$. Note that the energy
gap (energy at zero momentum) of the Higgs $\Delta$ is larger than
the one predicted by just assuming a change in the sign of the
mass squared coefficient. If phenomenologically needed it is
possible to add a mass, small with respect to the chemical
potential, for the $\eta$ field. Such a mass would induce a small
breaking of the $U_A(1)$ of the type:
\begin{eqnarray}
{{\cal L}_{m_{\eta}}}= - \frac{m^2_{\eta}}{2} \eta^2 \ .
\end{eqnarray}
The new gaps for $h$ and $\eta$ are:
\begin{eqnarray}
\Delta^2 &=& 6\mu_A^2\,\left[1 + \frac{1}{9}\frac{m^2_{\eta}}{\mu_A^2} +
O\left( \frac{m^4_{\eta}}{81\mu_A^4}\right) \right] \ , \nonumber \\
\Delta^2_{\eta} &=& \frac{2}{3}m^2_{\eta}\, \left[1  -
\frac{1}{9}\frac{m^2_{\eta}}{\mu_A^2} +
O\left( \frac{m^4_{\eta}}{81\mu_A^4}\right) \right] \ , \nonumber \\
\end{eqnarray}
where we have set $m=0$. Interestingly the corrections to the gaps
due to the $U_A(1)$ breaking term are further suppressed due to an
extra factor of $9$ appearing in the argument of the power series
expansion in $m^2_{\eta}/9\mu_A^2$. In the following we study the
theory with intact $U_A(1)$ and hence set $m_{\eta}=0$.

\section{The non Higgs Sector}\label{quattro}
We have shown that, at the tree level, the gauge bosons acquire
the ordinary electroweak masses and dispersion relations while we
argued that deviations with respect to the ordinary Higgs
mechanism, in our theory, arise when considering higher order
corrections. In this section we analyze some of the effects of
spontaneous symmetry breaking via a non zero $U_A(1)$ charge
density on the non Higgs sector of the electroweak theory due to
such higher order effects. We first investigate the gauge boson
sector and then the fermion one. On general grounds we expect the
presence of the chemical potential to induce different time and
spatial corrections while keeping rotational invariance intact.

\subsection{The Gauge Bosons}
The Higgs propagator is modified in the presence of the chemical
potential and assumes the form:
\begin{eqnarray}
\parbox{2cm}{\includegraphics[width=20cm,clip=true]{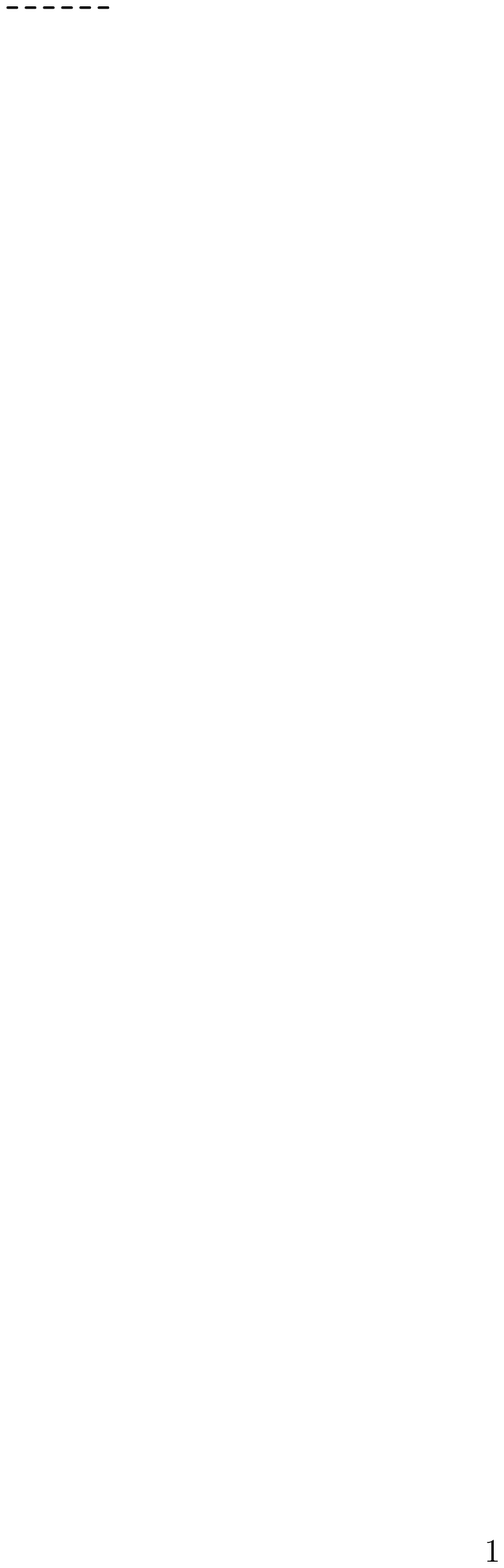}}
&=& i\,\frac{p^2}{p^4-2(\mu^2_A-m^2)p^2-4\mu^2_Ap_0^2} \ .
\end{eqnarray}
All of the loops containing this propagator are affected by the
presence of the chemical potential. The Landau gauge
\cite{Cheng-Li} is chosen to evaluate the relevant contributions
although our results are gauge independent. To set the conventions
the vacuum polarizations are defined as:
\begin{eqnarray}
i\,\Pi_{XY}^{\mu\nu}(q^2)=\int\,d^4x\,e^{-i\,q\cdot x} \langle
J^{\mu}_X(x)\,J^{\nu}_Y (0) \rangle \ ,\end{eqnarray} where $XY$
stands for the gauge boson indices $11,22,33,3Q$ and $QQ$
\cite{Peskin:1991sw}. The new contributions to the vacuum
polarizations are deduced by dividing $\Pi_{XY}^{\mu\nu}(q^2)$
into two parts \cite{Takeuchi:1992bu}:
\begin{eqnarray}
\Pi_{XY}^{\mu\nu}(q^2)={\Pi_{XY}^{\mu\nu}}^{SM}(q^2)
+{\Pi_{XY}^{\mu\nu}}^{new}(q^2) \ .
\end{eqnarray}
The first term is the contribution of the Standard Model physics
to $\Pi_{XY}^{\mu\nu}$ and the second term is the contribution of
the new Higgs sector physics. We are interested in computing the
new physics corrections for the $W$ vacuum polarization due to a
different Higgs sector with respect to the conventional Standard
Model one. The diagrams needed are
\cite{Peskin:1991sw,Takeuchi:1992bu}:
\begin{equation}
\parbox{9cm}{\includegraphics[width=9cm,clip=true]{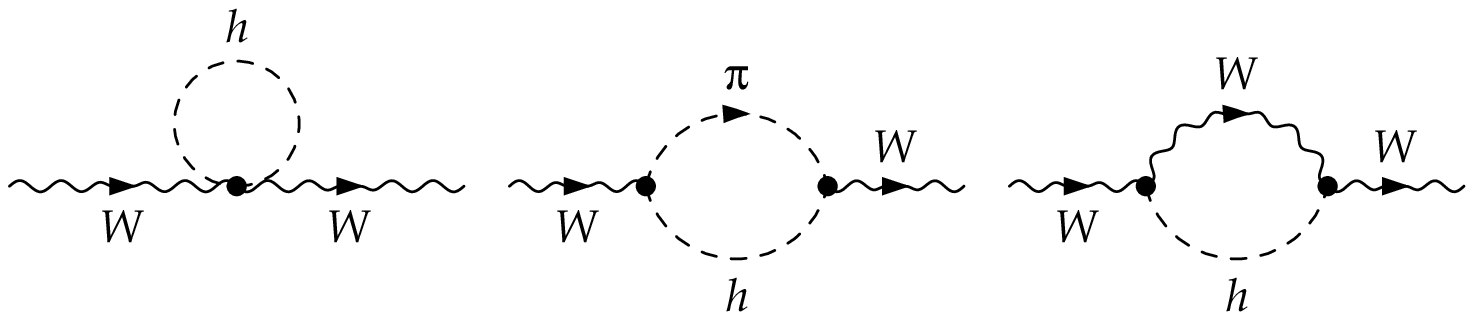}} \label{WZ-Self-Energies}
\end{equation}
The major difference with respect to known possible extensions is
the appearance of a new type of dispersion relations for the
Higgs. The $\eta$ particle does not appear in the previous
diagrams since we used a polar decomposition for the $M$ field. In
the appendix we explicitly compute the diagrams while here we
report the results for the vacuum polarizations, in the leading
logarithmic approximation\footnote{The finite contributions are
shown in the appendix.}, when the external momentum vanishes and
assuming an expansion in the gauge bosons masses with respect to
the Higgs mass. We also
 set without loss of generality $m=0$, the result is:
\begin{eqnarray}
\Pi_{WW}^{\mu\nu}(0)&=&
\frac{3\,g^2}{4\,(4\pi)^2}\,\log\left(\frac{\mu^2_A(2+\sqrt{3})}{\Lambda^2}\right)
\left[g_{\mu\nu} \,\left( M^2_W +
\frac{\mu^2_A}{9}\right)-\frac{4\,\mu^2_A}{9}\,V_{\mu}V_{\nu}
\right] + {\cal O}\left(\frac{M^2_{W}}{\mu^2_A}\right)\ ,
\end{eqnarray}
where $\Lambda$ is the renormalization scale and
$V_{\mu}=\left(1,\bf{0}\right)$. The photon vacuum polarization is
not affected at zero momentum due to the ordinary Ward identities
\cite{Takeuchi:1992bu}. The first diagram on the right hand side
of eq.~(\ref{WZ-Self-Energies}) does not spoil the Lorentz
covariance of the vacuum polarization which is due to the second
and third diagrams. Note that the specific combination
$\mu_A^2\left(2+\sqrt{3}\right)$ appearing in any logarithmic
corrections is consequence of the fact that in the presence of the
chemical potential the particle gaps are not the curvatures
evaluated on the minimum.

It is interesting to directly compare our results with the same
vacuum polarizations at zero momentum predicted in the Standard
Model \cite{Kennedy:1988sn}:
\begin{eqnarray}
{\Pi_{WW}^{\mu\nu}}^{SM}(0)&=&
\frac{3g^2}{4\,(4\pi)^2}\,g_{\mu\nu}\,
 M^2_W\,\log\left(\frac{M^2_H}{\Lambda^2}\right)+{\cal O}\left(\frac{M^2_{W}}{M^2_H}\right)\ .
\end{eqnarray}
$M_H$ is the conventional Higgs mass corresponding to the
curvature of the Higgs potential evaluated at the minimum i.e.
$2\mu^2_A$ (for $m=0$). The contribution to the $Z$ vacuum
polarization is obtained by replacing in the previous expressions
$g^2$ with $g^2/\cos^2\theta_W$ and $M^2_W$ with $M^2_Z$. The
onset of Lorentz breaking in this sector is small especially if
one chooses $\Lambda$ of the order of $\mu_A$. In general it is
possible to define a new set of oblique parameters capable to
capture the relevant corrections due to this type of spontaneous
symmetry breaking. Here, for illustration, we consider the
following straightforward extension of the parameter which
measures deviations with respect to the breaking of the custodial
symmetry i.e. the parameter $T$ \cite{Peskin:1991sw}:
\begin{eqnarray}
\alpha\,T^{\mu\nu}=\frac{e^2}{\sin^2\theta_W\,\cos^2_W\,M^2_Z}
\left({\Pi_{11}^{\mu\nu}}^{new}(0)-{\Pi_{33}^{\mu\nu}}^{new}(0)\right)
\ ,\end{eqnarray} with $\alpha=e^2/4\pi$ the fine structure
constant. We also used
$g^2\,\Pi_{11}^{\mu\nu}(0)=\Pi_{WW}^{\mu\nu}(0)$ and
${g^2}\Pi_{33}^{\mu\nu}(0)=\cos^2\theta_W\Pi_{ZZ}^{\mu\nu}(0)$
since the photon vacuum polarization at zero momentum vanishes.

This parameter is equal to $\alpha\,T\,g^{\mu\nu}$ for any Lorentz
preserving extension of the Higgs sector. The newly defined
parameter is not directly a measure of the amount of Lorentz
breaking but rather it estimates the amount of custodial symmetry
breaking for the different space time components of the vacuum
polarizations. Here we still have $T^{\mu\nu}=T\,g^{\mu\nu}$ but
due to the fact that we have different dispersion relations and a
different gap structure with the respect to the Standard Model
masses $T$, although very small, is not zero:
\begin{eqnarray}
T\approx-\frac{3}{16\pi}\frac{1}{\cos^2\theta_W}\log\left(1+\frac{\sqrt{3}}{2}\right)\approx
-0.048 \ .
\end{eqnarray}
Here we assumed the Standard Model Higgs mass to be given by the
expression $M^2_H=2\mu_A^2$ which is the curvature evaluated on
the minimum.
\subsection{The Fermions}
The fermions constitute a very interesting sector to be explored
since it can be used experimentally to test the idea presented in
this paper. In order to understand the type of corrections we
start with recalling that the chemical potential explicitly breaks
$SL(2,\mathbb{C})$ to $SO(3)$. So the corrections must
differentiate time from space in the fermion kinetic term
according to:
\begin{eqnarray}
(1 - a_0)\,\bar{f}\gamma^0\partial_0\, f+ (1-a)
\,\bar{f}\gamma^{i}\partial_i\, f\rightarrow
\bar{f}\gamma^0\partial_0\, f + v_{f}\,
\bar{f}\gamma^i\partial_i\, f \ , \qquad {\rm with} \qquad
v_{f}\simeq 1-(a-a_0) \ , \label{fermion-dr}
\end{eqnarray}
where $a$ and $a_0$ are the corrections induced by loop
contributions and $f$ represents a generic Standard Model fermion.
In the last expression we have rescaled the fermion wave function
and used the fact that the $a$'s are small calculable corrections.
In the difference $a-a_0$ all of the Lorentz covariant corrections
disappear while only the Lorentz breaking terms survive.

In order for the fermions to receive one loop corrections
sensitive to the chemical potential they need to couple at the
tree level with the Higgs. This is achieved via the Yukawa
interactions:
\begin{eqnarray}
\widetilde{Y}_{f}\, h \bar{f}f \ , \qquad {\rm with} \qquad
\widetilde{Y}_{f}\simeq \frac{m_{f}}{v} \ .
\end{eqnarray}
To determine $v_{f}$ at the one loop we need to compute only the
contributions to the fermion self energy which break Lorentz
invariance. In the present case the one loop diagram is
\begin{equation}
\parbox{3.5cm}{\includegraphics[width=3.5cm,clip=true]{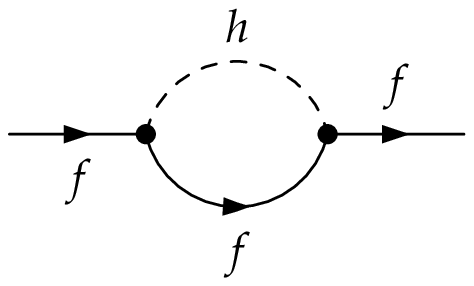}}
 \label{ftext_integral}
\end{equation}
\vskip.5cm where the solid line represent the fermion and the
dashed the Higgs field. The diagram is evaluated in detail in the
appendix and yields the following contribution to the $a$ and
$a_0$ coefficients:
\begin{eqnarray}
a&=&\frac{\widetilde{Y}_{f}^2}{2(4 \pi)^2} \left[\log
\left(\frac{\mu^2_A(2+\sqrt{3})}{\Lambda^2}\right)-
\frac{1}{6}\right] \ , \\
a_0&=&\frac{\widetilde{Y}_{f}^2}{2(4 \pi)^2}
\left[\log\left(\frac{\mu^2_A(2+\sqrt{3})}{\Lambda^2}\right)+4\sqrt{3}
-\frac{15}{2}\right] \ ,
\end{eqnarray}
If only the leading logarithmic corrections are kept we have that
the fermions still obey standard Lorentz covariant dispersion
relations. The dispersion relations are modified when considering
the finite contributions. To estimate the size of possible
departures from the standard dispersion relations we keep the
constant terms and determine
\begin{eqnarray}
v_{e}\simeq 1 - \widetilde{Y}_e^2\frac{0.4}{2(4\pi)^2} \simeq 1 -
5\times 10^{-15} \ ,
\end{eqnarray}
where the numerical evaluation has been performed for the
electron. The present formula is valid practically for all of the
fermions. For the muon the corrections to its velocity are of the
order of $8\times 10^{-10}$. The induced corrections for the
photon due to fermion loops are further suppressed by powers of
the fine structure constant $\alpha=e^2/4\pi$.

We briefly address how to construct the Yukawa terms for the
fermions in the present model. We assume the Standard Model
fermions not to carry the $U_A(1)$ charge since if they did they
would also couple directly to the chemical potential and hence
would develop a large Fermi surface. At this point there are
several possibilities to provide a mass to the fermions. {}It is
still possible to use the Standard Model term
\begin{eqnarray}
\widetilde{Y}_e\, \bar{E}_L\,M\frac{\left(1-\tau^3\right)}{2}E_R
\, + {\rm h.c.}= \widetilde{Y}_e \left(v + h \right)\, \bar{e}\,
e\ , \label{mass}
\end{eqnarray}
with
\begin{eqnarray}
E_L=\left(%
\begin{array}{c}
  \nu_L \\
  e_L \\
\end{array}%
\right)\ ,\qquad E_R=\left(%
\begin{array}{c}
  \nu_R \\
  e_R \\
\end{array}%
\right) \ . \end{eqnarray}
 Since the Yuakawa couplings are small
(except for the top quark) these terms would induce an explicit
but small $U_A(1)$ breaking. However, higher dimensional operators
preserving the $U_A(1)$ can be constructed:
\begin{eqnarray}
Y_e\, \bar{E}_L\,M\frac{\left(1-\tau^3\right)}{2}E_R \,\frac{{{\rm
det}M^{\dagger}}^{\frac{1}{2}}}{\widetilde{\Lambda}} + {\rm h.c.}=
\widetilde{Y}_e \left(v + h + \frac{h^2}{v} \right)\, \bar{e}\, e\
, \qquad {\rm with } \qquad \widetilde{Y}_e=Y_e
\frac{v}{\widetilde{\Lambda}}=\frac{m_e}{v} \ .\label{mass}
\end{eqnarray}
We used a doublet notation for the right fields for esthetic
reasons and $\widetilde{\Lambda}$ is an energy scale larger or of
the order of the electroweak scale $v$. {}For illustration we have
taken $Y_e$ to be real and $e$ is the electron. To provide a Dirac
mass to the upper component field we need to replace $1-\tau^3$
with $1+\tau^3$. In the last step of eq.~(\ref{mass}) we expanded
$M$ around the vacuum value in the unitary gauge and used the
polar decomposition of $M$. Any other fermion $f$ can acquire mass
in a similar way.

Other ways of providing a mass term to the fermions can be
explored. We expect, though, that the corrections to the fermion
dispersion relations induced by the Higgs sector under
consideration to be valid in general.

Another class of indirect corrections to the fermion sector are
the ones induced by modified gauge vector propagators discussed in
the previous section. To illustrate these effects we use the low
energy electroweak effective theory for the charged currents. The
neutral currents will be affected in a similar way. The chemical
potential leaves intact the rotational subgroups of the Lorentz
transformations, so the effective Lagrangian modifies as follows:
\begin{eqnarray}
{\cal L}^{\rm CC }_{\rm Eff}=-2\sqrt{2}\,G
J_{\mu}^{+}{J^{\mu}}^{-} \Rightarrow
-2\sqrt{2}\,G\,\left[{J_{\mu}}^{+}{J^{\mu}}^{-}+ \delta\,
J_{i}^{+}{J^{i}}^{-}\right] \ ,
\end{eqnarray}
where $\delta$ is a coefficient effectively measuring the
corrections due to modified dispersion relations for the gauge
vectors. Using the previous Lagrangian the decay rate for the
process $\mu\rightarrow e \bar{\nu}_{e}\nu_{\mu}$ is:
\begin{eqnarray}
\Gamma[\mu\rightarrow e \bar{\nu}_{e}\nu_{\mu}]=\frac{G^2
M_{\mu}^5}{192\pi^3}\left(1 + \frac{3}{2}\delta\right) \ ,
\end{eqnarray}
where $M_{\mu}$ is the muon mass and we neglected the electron
mass. However the effects of a non zero electron mass are as in
the Standard Model case \cite{Hagiwara:fs}. The parameter $\delta$
can be estimated using the vacuum polarizations presented in the
previous section yielding:
\begin{equation}
\delta\approx\frac{g^2}{3(4\pi)^2}\frac{\mu^2}{M_W^2}\log\left(\frac{(2+\sqrt
3)\mu^2}{\Lambda^2}\right).
\end{equation}
By choosing the renormalization scale $\Lambda$ to be of the order
of the electroweak scale $\sim M_Z$ and $\mu\simeq 150$~GeV we
determine $\delta\simeq 0.007$. Precise and independent
measurements of $M_W$ and the muon decay rate may observe
deviations with respect to the Standard Model. Finally we expect
sizable corrections to the fermion dispersion relations in
eq.~(\ref{fermion-dr}) induced by the gauge boson exchanges. These
arise in the fermion vacuum polarization at the two loop level and
are expected to be of the order of ${g^2}\delta/{(4\pi)^2}$.

\section{Discussion} \label{cinque}

We have studied some of the effects of a net background charge
associated to the global symmetry $U_A(1)$. When considering
possible non Abelian extensions of the Higgs sector such as the
ones naturally present in grand unified theories only chemical
potentials corresponding to mutually commuting charges can be
introduced. One also has to differentiate between chemical
potentials introduced for global and local symmetries. The main
difference relies on the fact that given a generic thermodynamic
potential $\Omega\left[T;\mu_{\rm G},\mu \right]$ we need to
impose the following constraints:
\begin{eqnarray}
\rho= -\frac{\partial \Omega[T;\mu_{\rm G},\mu]}{\partial \mu} \ ,
\qquad  0= \frac{\partial \Omega[T;\mu_{\rm G},\mu ]}{\partial
\mu_{\rm G}} \ ,
\end{eqnarray}
where $T$ is the temperature, $\mu$ is a generic chemical
potential associated to a globally conserved charge and $\mu_{\rm
G}$ is related to a gauge symmetry. These relations express the
fact that for charges associated with a global symmetry there is
no particular reason why the universe should be neutral. The
situation is more delicate for a gauge symmetry
\cite{Haber:1981ts} for which it is usually assumed that the
universe must be globally neutral. Actually this is strictly true
only if the universe is gravitationally closed. We however take,
following the literature, the second equation to hold. So, after
having determined the thermodynamical potential evaluated on the
physical vacuum of the theory one should determine the relations
between the various chemical potentials. In general, a non zero
$\mu_{\rm G}$ chemical potential is induced to insure gauge
neutrality. The gauge group $SU_L(2)\times U_Y(1)$ admits two
mutually commuting charges which are any linear combination of the
hypercharge and weak isospin \cite{Kapusta:aa}. It can be shown
that in the present case, at zero temperature and in the absence
of the chemical potential for the fermions
\cite{Kapusta:1990qc,{Gynther:2003za}}, electroweak neutrality is
guaranteed for a zero value of the chemical potentials associated
to the two commuting gauge charges and hence our results are
unaffected. We also stress that it is not possible to directly
compare the Higgs vacuum energy with the cosmological constant
since other contributions are supposed to be present in the
theory. Due to these other unknown contributions a large chemical
potential alone does not imply either a large charge or energy
density.

We have introduced a new mechanism able to explain the spontaneous
symmetry breaking in a generic gauge theory as due to the presence
of a net charge density background in the universe. The present
mechanism requires the Higgs field to posses global and gauged
internal symmetries. To some of the globally conserved charges
commuting with all of the gauge transformations we associate a non
zero chemical potential. This induces a negative type mass squared
triggering spontaneous symmetry breaking of the global and local
gauge transformations. We have studied the effects on the
electroweak sector of the Standard Model as a relevant example in
which the Higgs sector is minimally extended to have an extra
global $U_A(1)$ symmetry with a non zero  chemical potential. The
Bose-Einstein condensation of the Higgs induces new type of tree
level dispersion relations specific for the Higgs field while the
dispersion relations of the gauge bosons and fermions remain
unmodified. The latter are affected when considering higher order
corrections involving the Higgs field. To show how this occurs we
explicitly evaluated them for the vacuum polarizations of the
gauge bosons at zero external momentum. These corrections, due
entirely to the new Higgs sector of the theory, are known to
affect some of the oblique parameters. We first suggested how to
generalize the parameter $T$ to be applicable to our new Higgs
sector and then have shown that the value of $T$ is tiny and
negative in sign. This is so since our model preserves the
custodial symmetry. What is interesting is that $T$ is sensitive
to the new way of breaking the symmetry.

The fermion sector of the Standard Model is, on general grounds,
affected by the new mechanism via higher order corrections. We
expect for a generic fermion the appearance of modified dispersion
relations of the type $E^2 = v_{f}^2\,p^2+m_f^2$. At the one loop
level due to the smallness of the Yukawa coupling $v_f$ is small.

Spontaneous breaking of a gauge theory via Bose-Einstein
condensation necessarily introduces Lorentz breaking since a frame
must be specified differentiating time from space.  This must be
contrasted with different type of Lorentz violation due to an
intrinsically modified gravitational theory. The advantage with
our proposal is that it leads to distinctive and computable
corrections.

We recall that the issue of Lorentz breaking has recently
attracted much theoretical
\cite{Kostelecky:2002hh,Carroll:2001ws,{Amelino-Camelia:2002dx}}
and experimental attention \cite{Carroll:vb}. For example in
\cite{Carroll:2001ws} possible extensions of the Standard Model
were studied in which Lorentz symmetry is violated. In
\cite{Kostelecky:2002hh,Carroll:2001ws} the authors investigate in
some detail theoretical extensions and possible observational
consequences of the Standard Model in the presence of Lorentz
breaking. Here the underlying gravitational theory is not the
cause of Lorentz breaking which is due instead to having immersed
the theory in a background charge.

Many new avenues are left unexplored. {}For example, one can
imagine different types of Higgs scenarios where the chemical
potential effects and hence the Lorentz breaking is more or less
felt by the Standard Model particles. A very interesting
possibility would be to study supersymmetric extensions of our
model. Since fermions and bosons have different statistics the
chemical potential would also behave as an explicit source of
supersymmetry breaking. Cosmological consequences are currently
under exploration.

\acknowledgments We are very happy to thank P.H. Damgaard, H.B.
Nielsen and J. Schechter for discussions and comments. P.H.
Damgaard  and J. Schechter also for encouragements and a critical
reading of the manuscript. We acknowledge discussions with S.
Dawson, S. Hannestad, A. M\'{o}csy and P. Olesen. The work of F.S.
is supported by the Marie--Curie fellowship under contract
MCFI-2001-00181.

\appendix

\section{Evaluating Feynman Diagrams}

In this section we evaluate some of the Feynman diagrams. The set
of diagrams needed to estimate the corrections due to our modified
Higgs sector to the $W$ vacuum polarization at the one loop order
is:
\begin{equation}
\parbox{9cm}{\includegraphics[width=9cm,clip=true]{MWFullEPS.eps}}
\end{equation}
The dashed (dotted) line in the following equations stand for $h$
($\eta$) propagator while the dashed-dotted line connects an $h$
to an $\eta$ field:
\begin{eqnarray}
\parbox{2cm}{\includegraphics[width=20cm,clip=true]{hprop.ps}}
&=& i\,\frac{p^2}{p^4-2(\mu^2_A-m^2)p^2-4\mu^2_A\,p_0^2} \\
\parbox{2cm}{\includegraphics[width=20cm,clip=true]{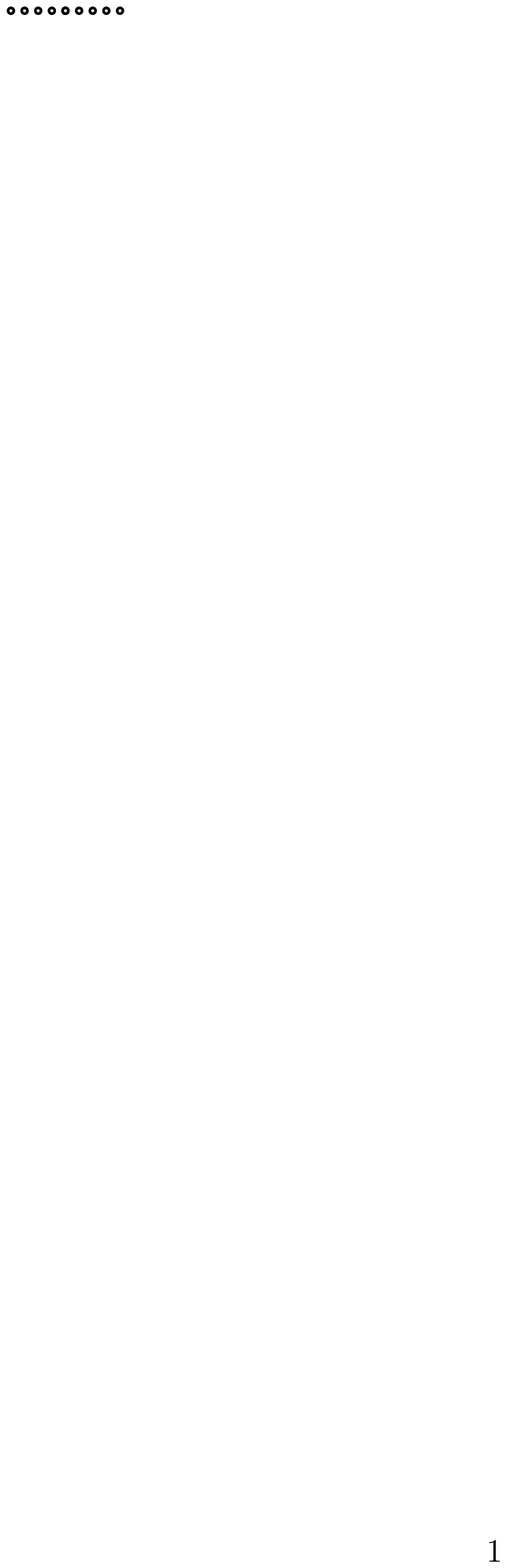}}
&=& i\, \frac{p^2 -
2(\mu^2_A-m^2)}{p^4-2(\mu^2_A-m^2)p^2-4\mu^2_A\,p_0^2} \\
\parbox{2cm}{\includegraphics[width=20cm,clip=true]{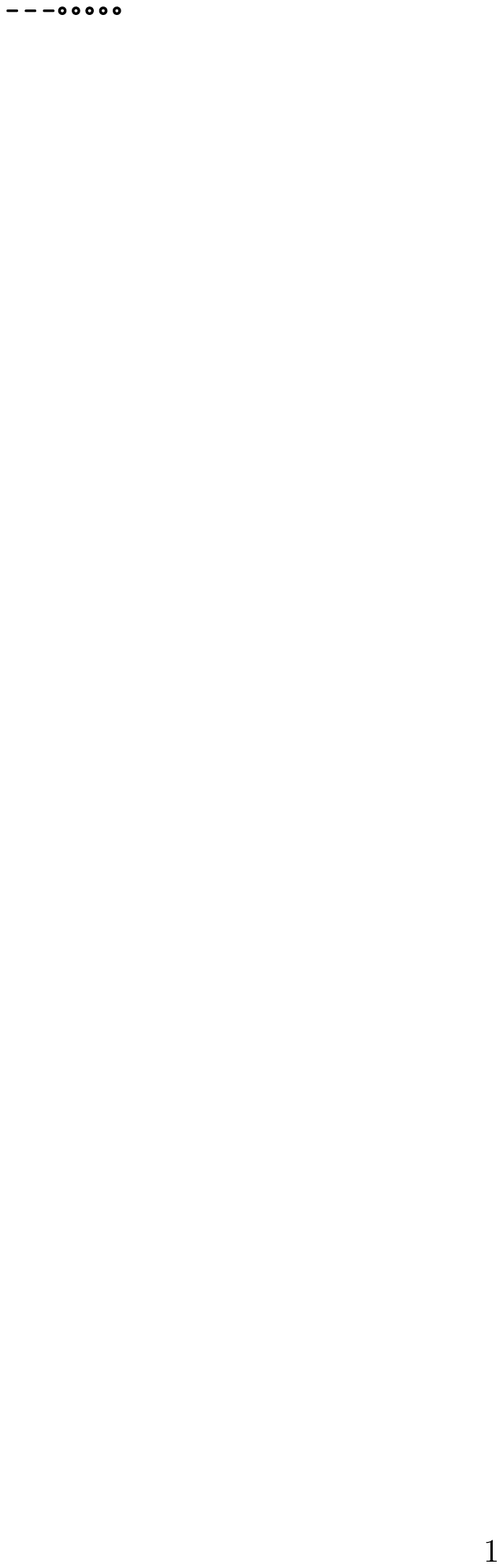}}
&=& \frac{2\,p_0 \mu_A}{p^4-2(\mu^2_A-m^2)p^2-4\mu^2_A\,p_0^2}
\end{eqnarray}
In the last line we have a negative sign if we have a
dotted-dashed line instead.

We now evaluate in some detail the following diagram where without
any loss of generality we set $m=0$.
\begin{eqnarray}
\parbox{3cm}{\includegraphics[width=3cm,clip=true,trim=0 -42 0 0]{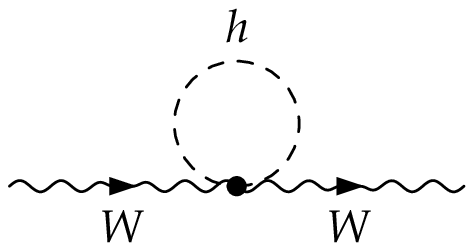}}
&=& i\frac{g^2g_{\mu\nu}}{4}
\,\,\int\frac{d^4p}{(2\pi)^4}\frac{p^2}{p^4+2p^2\mu^2_A+4\mu^2_Ap_0^2}\nonumber \\
&=& i\frac{g^2g_{\mu\nu}}{4}\frac{4\pi}{(2\pi)^4}\int_0^\infty dp
\,p^3\,\int_0^\pi
d\theta\frac{\sin^2\theta}{p^2+2\mu^2_A+4\mu^2_A\cos^2\theta}\nonumber \\
&=&
i\frac{g^2g_{\mu\nu}}{4\mu^2_A}\frac{1}{(4\pi)^2}\,\int_0^\infty
dp \,p^3\,\left[\sqrt{\frac{p^2+6\mu^2_A}{p^2+2\mu^2_A}}-1\right]\nonumber \\
&=& i
\frac{g^2}{4\left(4\pi\right)^2}\,g_{\mu\nu}\left[{\Lambda^2}+3\,{\mu^2_A}\left(
\log\left(\frac{\left(2+\sqrt{3}\right)\mu^2_A}{\Lambda^2}\right)-\frac{1}{6}\right)\right]+O(1/\Lambda)\
.
\end{eqnarray} We euclideanized the time
$p_0\rightarrow i\,p_0$, used the spherical coordinates and
denoted an ultraviolet cutoff by $\Lambda$. The previous diagram
is independent of the external momentum and is quadratically
divergent. This divergence is identical to the one within the
Standard Model. When adding together the other contributions the
quadratic divergence disappears as expected. The other diagrams
depend on the external momentum and we evaluated them in the zero
external momentum limit. The following diagrams are evaluated in
the Landau gauge \cite{Cheng-Li}
\begin{eqnarray}
\parbox{3cm}{\includegraphics[width=3cm,clip=true]{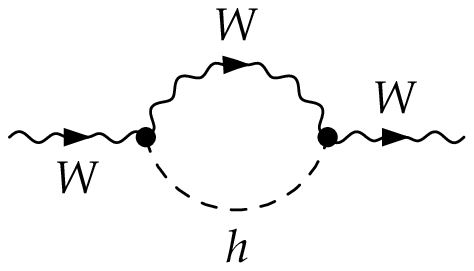}}
&=&
g^2M_W^2\int\frac{d^4p}{(2\pi)^4}\frac{p^2}{(p^2-M_W^2)(p^4-2\mu^2_A
p^2-4\mu^2_Ap_0^2)}\left[-g_{\mu\nu}+\frac{p_\mu
p_\nu}{p^2}\right]\nonumber \\ &\equiv & g_{\mu\nu}{\cal{I}}+{\cal
I}_{\mu\nu},
\end{eqnarray}
where we find
\begin{equation}
{\cal{I}}=i\frac{g^2}{\left(4\pi\right)^2}{M_W^2}\,\left[
\log\left(\frac{(2+\sqrt 3)\mu^2_A}{\Lambda^2}\right) +\sqrt{3} -
2\right] + {\cal O}\left(\frac{M^2_{W}}{\mu^2_A}\right)\ .
\end{equation}
 ${\cal{I}}_{\mu\nu}$ has non vanishing components only for $\mu=\nu$:
\begin{eqnarray}
{\cal{I}}_{00}
&=&-i\frac{g^2}{4\left(4\pi\right)^2}\,M^2_W\,\left[\log\left(\frac{(2+\sqrt
3)\mu^2_A}{\Lambda^2}\right)-2\sqrt{3}+\frac{7}{2}\right] + {\cal
O}\left(\frac{M^2_{W}}{\mu^2_A}\right) \ ,
\end{eqnarray}
and
\begin{eqnarray}
{\cal{I}}_{11} &=&i\frac{g^2}
{4\left(4\pi\right)^2}\,M^2_W\,\left[\log\left(\frac{(2+\sqrt
3)\mu^2_A}{\Lambda^2}\right)+2\sqrt{3}-\frac{23}{6}\right] + {\cal
O}\left(\frac{M^2_{W}}{\mu^2_A}\right)\nonumber \\ \nonumber \\&=&
{\cal{I}}_{22}={\cal{I}}_{33} \ .
\end{eqnarray}
The next diagram completes the vacuum polarization:
\begin{eqnarray}
\parbox{3cm}{\includegraphics[width=3cm,clip=true]{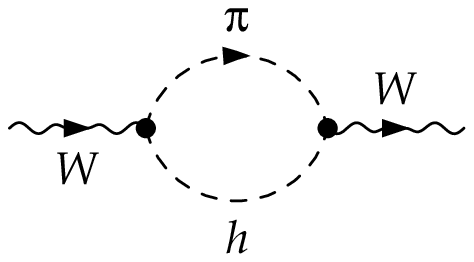}}
&=&
g^2\,\int\frac{d^4p}{(2\pi)^4}\frac{p_{\mu}p_{\nu}}{p^2}\frac{p^2}{p^4-2\mu^2_Ap^2-4\mu^2_Ap_0^2}
\equiv \widetilde{\cal I}_{\mu\nu} \ ,
\end{eqnarray}
where we find
\begin{eqnarray}
\widetilde{\cal{I}}_{00}
&=&-i\frac{g^2}{4\left(4\pi\right)^2}\,\left[\Lambda^2
+4\,\mu^2_A\left(\log\left(\frac{(2+\sqrt
3)\mu^2_A}{\Lambda^2}\right)-\sqrt{3}+\frac{11}{6}\right)\right]
 \ ,
\end{eqnarray}
and
\begin{eqnarray}
\widetilde{\cal{I}}_{11}
&=&i\frac{g^2}{4\left(4\pi\right)^2}\,\left[\Lambda^2
+\frac{8}{3}\,\mu^2_A\left(\log\left(\frac{(2+\sqrt
3)\mu^2_A}{\Lambda^2}\right)+\frac{\sqrt{3}}{2}-\frac{7}{6}\right)\right]\nonumber \\
\nonumber\\ &=& \widetilde{\cal{I}}_{22}=\widetilde{\cal{I}}_{33}
\ .
\end{eqnarray}
%

The corrections to a generic Standard Model fermion can be
determined by computing the following one loop diagram:
\begin{equation}
\parbox{3cm}{\includegraphics[width=3cm,clip=true]{MFHEPS.eps}}
=i\widetilde{Y}_f^2\int\frac{d^4k}{(2\pi)^4}\frac{1}{(k^2+2\mu^2_A+4\mu^2_A\cos^2\theta)}
\frac{(p-k)_\mu\bar{f}_p\gamma^\mu f_p+m_f\bar{f}_p
f_p}{((p-k)^2+m_f^2)}\equiv {\cal{F}}_\mu. \label{f_integral}
\end{equation}
Where the solid line denotes a fermion and $f_p$ is the associated
Dirac spinor. Note that the numerator contains an implicit factor
of $i$ for the $\mu=0$ component. We are using an Euclidean metric
and to compute the diagram we choose a frame in which the incoming
fermion has momentum $p=(p_0,P,0,0)$. Since we are interested in
the small (with respect to the Higgs scale) external momentum we
expand the fermion propagator as follows:
\begin{equation}
\frac{1}{k^2-2k\cdot P}\approx\frac{1}{k}\ \left
(\frac{1}{k}+\frac{2(\cos\theta+\sin\theta\cos\phi)}{k^2}P\right
)+O(P^2)
\end{equation}
These limits allows us to explicitly perform all of the integrals
(\ref{f_integral}) leading to:
\begin{eqnarray}
{\cal{F}}_0&=&-iV\frac{p_0}{32\pi^2}\left
[-\frac{15}{2}+4\sqrt{3}+\ln{\frac{(2+\sqrt{3})\mu^2_A}{\Lambda^2}}\right
]+i\frac{m_f}{32\pi^2}\left
(4-2\sqrt 3-2\ln\frac{(2+\sqrt 3)\mu^2_A}{\Lambda^2}\right )\\
{\cal{F}}_1&=&iV\frac{P}{32\pi^2}\left
[-\frac{1}{6}+\ln{\frac{(2+\sqrt{3})\mu^2_A}{\Lambda^2}}\right
]+i\frac{m_f}{32\pi^2}\left
(4-2\sqrt 3-2\ln\frac{(2+\sqrt 3)\mu^2_A}{\Lambda^2}\right )\\
{\cal{F}}_2&=&{\cal{F}}_3=i\frac{m_f}{32\pi^2}\left (4-2\sqrt
3-2\ln\frac{(2+\sqrt 3)\mu^2_A}{\Lambda^2}\right ),
\end{eqnarray}
where $p_0=P$ if $m_f=0$ and $p_0=m_f$ for a massive heavy fermion
and used the tree level on shell relations. Other diagrams can be
computed following the outlined procedure.

\end{document}